\documentclass[1p,authoryear]{elsarticle}

\journal{Advances in Atomic, Molecular, and Optical Physics}

\usepackage{amssymb,amsmath}

\def\be{\begin{eqnarray}}
\def\ee{\end{eqnarray}}
\def\expect#1{\left\langle#1\right\rangle}
\def\Xe#1{$^{#1}$Xe}

\begin{document}

\begin{frontmatter}

\title{Spin-Exchange Pumped NMR Gyros}

\author[uw]{Thad G. Walker}  
\author[ngc]{Michael S. Larsen}
\address[uw]{Department of Physics, University of Wisconsin-Madison, Madison, WI 53706}
\address[ngc]{Northrop Grumman-Advanced Concepts and Technologies, Woodland Hills,  CA 91367}

\begin{abstract}
We present the basic theory governing spin-exchange pumped NMR gyros.  We review the basic physics of spin-exchange collisions and relaxation as they pertain to precision NMR.  We present a simple model of operation as an NMR oscillator and use it to analyze the dynamic response and noise properties of the oscillator.  
  We discuss the primary systematic errors (differential alkali fields, quadrupole shifts, and offset drifts) that limit the bias stability, and discuss methods to minimize them.  We give with a brief overview of a practical implementation and performance of an NMR gyro built by Northrop-Grumman Corporation, and conclude with some comments about future prospects.
  \begin{keyword}
Gyros \sep Nuclear Magnetic Resonance \sep Spin-Exchange Optical Pumping 
  \end{keyword}
\end{abstract}

\end{frontmatter}

\tableofcontents

\section{Introduction}

	Nuclear magnetic resonance gyros (NMRGs) based on spin-exchange optical pumping of noble gases have been developed over several decades of largely industrial research, first at Litton Industries and more recently at Northrop Grumman Corporation (NGC).  The basic physics of the production and detection of nuclear magnetic resonance using hyperpolarized noble gases has been extensively studied, and off-shoots of NMRGs in physics laboratories have achieved some of the highest sensitivity frequency measurements to date.  In this paper we present a mostly self-contained discussion of the physics and operation of NMRGs of the Litton/NGC type.
	
	Although spin-exchange optical pumping (SEOP) of He-3 was first demonstrated in 1960 by Bouchiat, Carver, and Varnum \citep{Bouchiat1960}, little follow-up occurred in academic laboratories in the 1960s and 1970s.  During that period, Litton began investigating the use of SEOP for gyroscopic applications \citep{Grover79patent}.  This work included the first demonstrations of SEOP of Ne, Kr, and Xe, the recognition of the remarkably high polarizations attainable, and the further enhancement of hyperpolarized NMR signals using in-situ magnetometry \citep{Grover78}.  With the advent of ring-laser and fiber-optic gyros, this work was discontinued at Litton in the mid 1980s.  Meanwhile,  W. Happer and his group at Princeton published an extensive set of investigations into the fundamental physics of hyperpolarized noble gases \citep{Happer84,Zeng85,WalkerRMP}.  This work led to the development of magnetic resonance imaging using hyperpolarized He and Xe \citep{Albert94,Middleton95}, high-density spin-polarized targets for nuclear and high-energy physics \citep{Singh2015}, neutron polarizers and analysers \citep{Chen2014}, extensive use of hyperpolarized Xe in chemical physics and NMR spectroscopy\citep{Ledbetter2012}, and further development of hyperpolarized gases for ultrasensitive spectroscopy in devices such as noble-gas masers \citep{Glenday2008,Rosenberry2001}, gyros \citep{kornack2005,Fang2013} and co-magnetometers for studies of fundamental symmetries \citep{Smiciklas2011,Brown2010}.
	
It is remarkable that with this tremendous range of applications of hyperpolarized noble gases, the original stimulating ideas from the 1970s about their use for NMRGs was never published beyond a single overview paper by  \citet{Kanegsberg1978}, a review by \citet{Karwacki1980}, patents, conference proceedings, and project reports to funding agencies.  This work was reviewed from a current perspective by \citet{Donley2010}. Interest in NMR gyros  revived in the early 2000s at NGC when it was realized that NMRGs have the potential to outperform other types of gyros for small, low-power applications.   This development is continuing \citep{Larsen2012,Meyer2014}, and an overview of the basic concepts of NMRGs was recently published by \citet{Donley2013}.  
The authors feel that it is timely to present a more detailed treatment of the physics of  spin-exchange-pumped NMRGs, in particular as implemented in the Litton/NGC design.  A parallel development has begun in China \citep{Liu2015}, and a related approach with applications to Xe edm searches is being pursued in Japan \citep{Yoshimi2008}.  Although the individual components of NMRGs have been studied in other contexts, the realization of hyperpolarized gas techniques into a small physical package with remarkable capabilities vis-a-vis sensitivity, accuracy, while simultaneously maintaining an impressive bandwidth are of considerable current research interest. Furthermore, new approaches are now being investigated (at Wisconsin and elsewhere), and an appreciation for the successes and challenges of the NGC NMRG are essential for proper evaluation of those new approaches.
	
This paper is organized as follows.  We begin with an overview of basic spin-exchange and NMR physics of importance to NMRGs, including a basic description of the physical implementation of an NMRG. We then present a simplified analysis of the operation of a single-species NMR oscillator that will elucidate the basic operation of an NMRG.  This naturally leads to a more sophisticated feedback analysis that will allow us to discuss issues such as scale factor, bandwidth, fundamental noise and systematic errors.  The latter include a simplified model of electric quadrupole effects, and a discussion of the``isotope effect" of the alkali field.  Dual isotope operation is then added, including a discussion of the suppression of clock phase noise when properly configured. We conclude with a discussion of the performance of a recent version of the NGC NMRG, and present some basic ideas concerning scaling of NMRGs.

\section{NMR Using Hyperpolarized Gases}

\subsection{Precession of nuclei due to magnetic fields and rotations}

The primary fundamental interaction between nuclear spins and their environment is through magnetic fields. { }In a stationary inertial frame
the energy of a nuclear spin \(\pmb{K}\) in a magnetic field \(\pmb{B}\) is \(H=-\hbar  \gamma  \pmb{B\cdot K}\) where the gyromagnetic ratio \(\gamma\)
is positive for Xe-131 and negative for all the other stable noble-gas isotopes. { }According to Ehrenfest{'}s Theorem, the time evolution of \(\langle
\pmb{K}\rangle\) is
\begin{equation}
\frac{d\langle \pmb{K}\rangle }{dt}=\frac{-i}{\hbar }\left\langle[\pmb{K},H]\right\rangle=i \gamma  \langle [\pmb{K,K\cdot B}]\rangle =\pmb{-}\gamma  \pmb{B}\pmb{\times }\langle
\pmb{K}\rangle
\end{equation}
which is the classical equation for the precession of a magnet in a magnetic field, generally called the Bloch equation in the NMR literature. {
}In what follows, we shall drop the expectation value symbols. 

In a uniform magnetic field \(\pmb{B}=B_z\hat{z}\) it is useful to focus on the nuclear spin-components parallel and perpendicular to the
magnetic field, \(\pmb{K}=K_z\hat{z}+\pmb{K}_\perp \). { }It is further convenient to use a phasor representation of \(\pmb{K}_\perp \),
defining \(K_+=K_x+i K_y=K_\perp e^{-i \phi }\). { }Then the Bloch equation becomes
\begin{equation}
\frac{dK_+}{dt}=-i\gamma  B_zK_+
\end{equation}
with solution
\begin{equation}
K_+(t)=K_\perp e^{-i \gamma  \int B_z \, dt}
\end{equation}
with a phase \(\phi =\gamma \int B_z \, dt\). 
Suppose the precession is detecting by measuring the component \(\hat{d}\cdot \pmb{K}_\perp \), where \(\hat{d}\) makes an angle
$\alpha $ with the x-axis.
\begin{equation}
\hat{d}\cdot \pmb{K}_\perp =K_\perp \cos (\phi +\alpha )
\end{equation}
The measurement device is fixed relative to the apparatus. { }If the apparatus is rotating about the $\hat{z}$-axis at an instantaneous frequency \(\omega _r=d\alpha /dt\),
the detected quantity is \(\phi +\alpha =\int \left(\gamma  B_z+\omega _r\right) \, dt\). { }Thus rotation is equivalent to a magnetic field \(\left.\omega
_r\right/\gamma\) and increases the Larmor frequency for Xe-131 while decreasing it for Xe-129 or He-3. { }This is equivalent to having the effective Hamiltonian
for the nuclei be
\begin{equation}
H=-\hbar  \left(\gamma  \pmb{B}+\pmb{\omega }_r\right)\cdot \pmb{K}
\end{equation}
For magnetometry applications, one would generally wish to pick large gyromagnetic ratios, while rotations will be generally easier to measure for
nuclei with small gyromagnetic ratios. { }Later in this paper we will discuss using dual species strategies to effectively eliminate either magnetic
or rotation sensitivities. 

An NMR instrument can also be used to search for exotic new physics and various versions of spin-exchange pumped nuclear-magnetic resonators
have been developed to do this. { }Examples include searches for electric dipole moments, violations of Lorentz invariance, and searches for scalar-pseudoscalar
couplings \citep{Glenday2008,Rosenberry2001,kornack2005,Smiciklas2011,Brown2010}.  Most of these experiments, while using NMR in various ways, are significantly different than the approach treated here and we encourage
interested readers to study the references for more information.   

   \subsection{A Minimal Spin-Exchange NMRG}

Figure 1 shows a basic spin-exchange NMR apparatus. Rubidium and isotopically enriched Xe, along with N$_2$ and H$_2$ buffer gases, are contained in a coated glass cell typically a few mm in size.  The Rb atoms are optically pumped with circularly polarized light propagating parallel to a magnetic field \(B_z\hat{z}\) that
defines the sensitive rotation axis for the gyro. { }The spin-polarized Rb atoms undergo collisions with Xe atoms. { }During these collisions, hyperfine interactions
\begin{figure}[htbp]
   \centering
   \includegraphics[width=3.0 in]{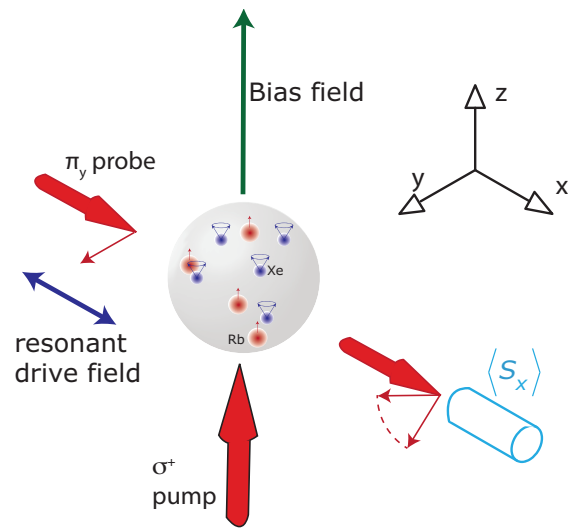} 
   \caption{Simple NMR gyro apparatus.  Rb atoms are spin-polarized by the pump laser, transfer angular momentum to the Xe nuclei via collisions, and detect the magnetic field produced by the precessing Xe nuclei by causing a Faraday rotation of the polarization of the probe laser.  The polarized Xe nuclei are driven to precess by the resonant oscillating drive field.  The phase shift between the drive field and the oscillation of the nuclear precession about the bias field direction changes when the apparatus rotates about the bias field direction.}
   \label{fig:example}
\end{figure}
between the Rb atoms and the Xe nuclei slowly transfer polarization to the Xe nuclei. { }The Xe nuclei reach a steady-state polarization of typically
10$\%$ after 10s of seconds of spin-exchange collisions with polarized Rb. { }Once polarized, the Xe nuclei can be induced to precess by applying
a transverse {``}drive{''} magnetic field that oscillates at a frequency near the Xe resonance frequency. { }The resonant drive tips the Xe nuclei
partially into the x-y plane, so the Xe nuclei then precess around the z-axis. { }The precessing Xe nuclei produce an oscillating y-magnetic field,
experienced by the Rb atoms, that causes the Rb atoms to tip slightly toward the x-axis. { }The resulting x-polarization of the Rb atoms produces
a different index of refraction for the $\sigma ^+$ and $\sigma ^-$ components of a linearly polarized probe laser that propagates along the
x-direction. { }The  rotation of the polarization of the probe light so produced is proportional to the y-component of the Rb spin polarization, and is thus a direct
measure of the precessing polarization of the Xe nuclei \citep{Lam1985patent}.  { }An electronic circuit filters the Xe signal, phase shifts it, and applies an amplified
version to the drive coils. { }This feedback loop ensures that the drive frequency is equal to the NMR resonance frequency. { }A frequency counter
registers the drive frequency. As long as the magnetic field is held steady, changes in the Larmor frequency are precisely equal to the rotation frequency of the apparatus.

\subsection{Spin-Exchange Optical Pumping}

The basic principles of spin-exchange optical pumping are well-known \citep{WalkerRMP} but they play an essential and non-trivial role in the physics of NMR gyros.
{ }Spin-exchange occurs due to the Fermi-contact hyperfine interaction between the alkali-metal atom and the noble-gas nuclei
\begin{equation}
H_{{se}}=\alpha (R) \pmb{S}\cdot \pmb{K}
\end{equation}
The coupling strength $\alpha(R)$ is proportional to the Rb spin-density at the position of the Xe nucleus, and thus depends strongly on the interatomic separation $R$.
The spin-exchange interaction has two primary effects. { }First, collisions of noble-gas nuclei with spin-polarized alkali atoms result in spin transfer from
the alkali electrons. { }These collisions are known to be of two types: collisions between atom pairs, and three-body collisions that form weakly
bound Rb-Xe van der Waals molecules. { }Both types of collisions are at work under typical NMRG conditions, and in  \ref{RbXe} we summarize the relevant formulas for RbXe spin-exchange, including providing numerical values suitable for estimates of spin-exchange collision rates under various conditions.
 
The second effect of the spin-exchange interaction is that the hyperfine interaction mimics an effective magnetic field that is proportional
to the alkali spin polarization, so that the Xe Larmor frequency is shifted by this {``}alkali field{''}. { }Likewise, the alkali atoms experience
an effective field proportional to the Xe nuclear polarization. { }These fields are conventionally compared in size to the magnetic field that would
be produced by a fictional uniform magnetization:
\begin{eqnarray}
 \pmb{B}_K&=&-\kappa \frac{8\pi  g_s\mu _B}{3}[A]\pmb{S}=b_{KS}\pmb{S}\nonumber \\
 \pmb{B}_S&=&\kappa \frac{8\pi  \mu _K}{3 K}[X]\pmb{K}=b_{SK}\pmb{K} 
\end{eqnarray}
Here $g_s\approx2$, $\mu_B$ and $\mu_K$ are the electron and nuclear magnetic moments, and $[A],[X]$ are the Rb and Xe densities.
The enhancement factor $\kappa $, about 500 for RbXe \citep{Ma2011}, arises from the close penetration of the alkali electron into the core of the noble gas, and
was one of the important discoveries in the early history of spin-exchange optical pumping \citep{Grover78,Schaefer89}. { }The enhancement of the noble-gas field \(B_K\)indicates
that the field detected by the alkali atoms is roughly 500 times larger than would be sensed by an NMR surface coil. { }This tremendous advantage
is somewhat offset by systematic effects of the alkali field \(B_S\) { }that need to be managed in a gyro application.

\subsection{Spin-relaxation of Polarized Noble Gases}

The gas-phase relaxation mechanisms for Xe nuclei are dominated by the spin-exchange collisions with the Rb atoms. These spin-exchange collisions
compete with spin-relaxation from diffusion through magnetic field gradients and in collisions with walls. { }A tremendous amount of effort has been
expended in devising walls with advantageous spin-relaxation properties. { }Generally, bare glass walls affect the spin-1/2 nuclei Xe-129 and He-3
quite minimally, so that relaxation times of minutes (Xe-129) to hours (He-3) are achievable with careful surface preparation. { }For nuclei with
electric quadrupole moments, such bare glass surfaces tend to have large electric field gradients that cause substantial relaxation. { }For such
nuclei, alkali-hydride coatings are advantagous and bring the wall-relaxation times for Xe-131 to tens of seconds for mm-scale
glass cells \cite{Kwon1981,Kwon1984patent,Kwon1984Bpatent,Kwon1984Cpatent} 

Magnetic field gradients are well-known to limit the transverse relaxation times for spin-exchange pumped nuclei.  Since NMR gyros will usually use magnetic shields to provide additional suppression of magnetic sensitivity, with field shimming it is usually possible for the transverse relaxation times to essentially reach the longitudinal relaxation time limit.

\subsection{Bloch Equations for Spin-Exchange Pumped NMR}

The net effect of rotations, spin-exchange collisions, the alkali field, and wall/magnetic-field-gradient relaxation are to modify the Bloch
equation to
\begin{equation}
\frac{d\pmb{K}}{dt}=-\left[\gamma  \left(\pmb{B}+b_{KS}\pmb{S}\right)+\pmb{\omega }_r\right]\times \pmb{K}+\Gamma _{{se}}(\pmb{S}-\pmb{K})-\tilde{\pmb{\Gamma
}}_w\pmb{\cdot }\pmb{K}
\end{equation}
where the relaxation matrix from wall collisions and magnetic field gradients is \(\tilde{\pmb{\Gamma }}_w\). { }Rather than explicitly separate
these effects from spin-exchange relaxation, it is convenient to lump them into a single relaxation matrix { }\(\tilde{\pmb{\Gamma }}=\Gamma
_2\left(\hat{x}\hat{x}+\hat{y}\hat{y}\right)+\Gamma _1\hat{z}\hat{z}.{  }\) Likewise, for much of our discussion the magnetic field, the
alkali field, and the rotation can be conveniently discussed as an effective Larmor frequency \(\pmb{\Omega }=\gamma  \left(\pmb{B}+b_{KS}\pmb{S})\right.+\pmb{\omega
}_r\). { }Then the Bloch equation becomes
\begin{equation}
\frac{d\pmb{K}}{dt}=-\pmb{\Omega }\times \pmb{K}-\pmb{\tilde{\Gamma }\cdot K}+\pmb{R}_{{se}}
\end{equation}
where \(\pmb{R}_{{se}}=\Gamma _{{se}}\pmb{S}\) is the spin-exchange pumping rate. { }The large Larmor frequency of the alkali atoms keeps
\(\pmb{S} \| \pmb{\Omega }\), so to a good approximation \(\pmb{R}_{{se}}\) is usually along the \(\hat{z}-\)axis.

\section{NMR Oscillator Basics}

In the following we analyze a simple model of the NMR Gyro.  We assume that the spin dynamics of the two Xe isotopes are well modeled by Bloch equations.  This is an excellent approximation for 129-Xe which is spin-1/2, but will ignore the quadrupole dynamics of 131-Xe.  In addition, the following treatment will ignore the isotope effect in the magnetic field of the Rb atoms.  We assume that there is a DC magnetic field applied along the z-axis and a feedback-generated oscillating magnetic field along the x-axis.  A more sophisticated model that accounts for the real spin-exchange and nuclear precession dynamics is being developed.	

The self-oscillation of the NMR Gyro can be understood by assuming that a transverse oscillating magnetic field is applied to the x-direction
of the gyro that is of constant amplitude and whose phase is delayed by an amount $\beta $ from the phase of the signal picked up along the y-direction.
{ }In other words, if the transverse coherence is \(K_+=K_x+i K_y=K_\perp e^{-i \phi }\), the Larmor frequency of the applied x-field is
\(-\Omega _d\sin [\phi -\beta ]\), with \(\Omega _d\) fixed in amplitude. { }The Bloch equations for the nuclear spin components are then
\begin{eqnarray}
 \frac{d K_+}{dt}&=&-\left(i \Omega _z+\Gamma _2\right)K_+-i \Omega _d\sin [\phi -\beta ]K_z \\
 \frac{d K_z}{dt}&=&-\Omega _d\sin [\phi -\beta ]\sin [\phi ]K_\perp -\Gamma _1 K_z+R_{{se}} 
\end{eqnarray}
The precession of the nuclei must also be supplemented by an electronic feedback network that drives the phase difference to a value \(\beta _0\)
which for various reasons may be chosen to be non-zero, thus running the oscillator somewhat off-resonance. { }In a first analysis, we will soon
assume that the feedback tightly locks the phase difference to the value \(\beta _0.\)

The amplitude and phase of the transverse polarization \(K_+\) obey quite different dynamics, thanks to the feedback. { }The real and imaginary
parts of Eq.$\, $ 10 lead to
\be
 \frac{d K_\perp }{dt}&=&-\Gamma _2K_\perp +\Omega _dK_z\sin [\phi -\beta ]\sin [\phi ]\nonumber \\ &\approx& -\Gamma _2K_\perp +\frac{\Omega
_dK_z}{2}\cos [\beta ]  \label{repart}\\
 \frac{d\phi }{dt}&=&\Omega _z+\frac{\Omega _dK_z}{K_\perp }\sin [\phi -\beta ]\cos [\phi ]\nonumber \\
 &\approx& \Omega _z-\frac{\Omega _dK_z}{2K_\perp }\sin[\beta ]  \label{impart}
\ee
The approximation made here is to neglect the small terms that oscillate at frequency 2\(\dot{\phi }\) (rotating wave approximation). { } Such terms
will quickly average to zero and will be neglected below. { }

Equation \ref{repart} gives a steady-state relationship between the transverse and longitudinal polarizations:
\be
 K_\perp=\frac{\Omega _dK_z}{2 \Gamma _2}\cos [\beta ]
\ee
which simplfies Eq. \ref{impart} to
\be
 \frac{d\phi }{dt}=\Omega _z-\Gamma _2\tan [\beta ] 
 \label{gyroeqn}
\ee
Notice that the transverse polarization does not depend on \(\Omega _{z }\), since $\beta$ is held constant.  The longitudinal polarization,
\be
 K_z&=&\frac{R_{{se}}}{\Gamma _1+\frac{\Omega _d^2}{4 \Gamma _2}\cos ^2[\beta ]} 
\ee
 is also independent of $\Omega_z$.  Thus these settle to their steady-state values, even if the Larmor frequency \(\Omega _z\) is varying in time:
which says that the spin is tipped away from the z-axis by an angle \(\tan [\Theta ]=\Omega _x\cos [\beta ]/2\Gamma _2\).\\

The gyro dynamics can now be understood by focusing on the fundamental gyro equation (\ref{impart}).  It can be rewritten as
\be
 \frac{d\phi }{dt}=\Omega _z-\Gamma _2\tan \left[\beta _0\right]=\Omega _z+\Delta  
 \label{gyroeqn2}
\ee
where \(\Delta =-\Gamma _2\tan \left[\beta _0\right]\) is the detuning off resonance. A key point to recognize is that as long as \(\beta _0\) is
held fixed, there is no damping term in (\ref{gyroeqn2}). { }The nuclear phase can change its precession rate fast compared to \(\Gamma _2\) and there are no significant polarization transients ($K_z$ and $K_{\perp}$ are unaffected). { }The nuclear phase is an accurate time-integral
of the Larmor precession frequency, and the bandwidth can greatly exceed \(\Gamma _2\).

\section{Detection of NMR Precession Using In-situ Magnetometry}

$\quad $As already noted, NMR detection in a spin-exchange NMRG is done using the alkali atoms as an integrated in-situ magnetometer. { }The EPR
frequency shift is greatly enhanced (a factor of 500 for Xe) by the Fermi contact interaction; the enhancement of the alkali electron density at
the site of the noble-gas nuclei produces an enhanced frequency shift.

$\quad $There are a variety of ways the integrated magnetometer could be configured, with various pros and cons. { }Generally, since the desired
signal is the transverse polarization of the noble gas, a vector magnetometer is preferred that is insensitive to \(B_z\) and maximally sensitive
to \(B_y=b_KK_y=b_KK_\perp \sin (\phi )\). { }A convenient method to accomplish this is to use parametric modulation \citep{Volk80b}. { }A sine wave oscillating
at the alkali Larmor frequency (100 kHz range) is applied along the \(\hat{z}-\)axis, and the electron spin develops \(S_x\) modulation at this frequency
in the presence of transverse polarization of the noble gas.

$\quad $A simplified treatment of the alkali magnetometer will be given here. { }Effects due to the alkali hyperfine structure, alkali-alkali spin-exchange
collisions, and the details of alkali relaxation will be ignored but careful consideration of these matters is necessary for actual implementation.
The Bloch equation for the alkali electron is
\be
 \frac{d S_+}{dt}=\left(i \left[\omega _0+\Omega _1\cos \left(\omega _1t\right)\right]-\Gamma _A\right)S_+-i \gamma _Ab_KK_+S_z 
\ee
where \(\omega _0\) is the alkali resonance frequency in the DC magnetic field, \(\Omega _1\cos \left[\omega _1t\right]\) is the applied parametric
modulation field and \(\Gamma _A\) is a phenomenological parameter describing the relaxation of the alkali spins. { }Moving to an {``}oscillating
frame{''}, \(S_+=A_+e^{i \mu _1}\), where \(\mu _1=\frac{\Omega _1}{\omega _1}\sin \left(\omega _1t\right)\), gives
\be
 \frac{d A_+}{dt}=\left(i \omega _0-\Gamma _A\right)A_+-i \gamma _Ab_KK_+S_ze^{-i \mu _1} \\
\ee
Assuming \(\omega _0>>\Gamma _A\), we can expand \(e^{i \mu _1}=J_0\left(\left| \mu _1\right| \right)+2 i J_1\left(\left| \mu _1\right|
\right)\sin \left(\omega _1t\right)+2 J_2\left(\left| \mu _1\right| \right)\cos \left(\omega _1t\right)\ldots\) to approximate
\be  A_+= \frac{-i \gamma _Ab_KK_+S_z}{\Gamma _A+i\left(\omega _1-\omega _0\right)}J_{-1}e^{i \omega _1t} 
\ee
Assuming detection of \(S_x\), the output of a lock-in demodulated with \(\cos \left[t \omega _1+\alpha \right]\) is
\be
\left\langle S_x\cos \left[\omega _1t+\alpha \right]\right\rangle &=& 
\frac{\gamma _Ab_K J_{-1} S_z}{2 \Gamma _A}\left({\sin}[\alpha ] \left(-J_0+J_2\right) K_x \right.\nonumber \\
&&+\left.{\cos}[\alpha ] \left(J_0+J_2\right) K_y\right)
 \ee
By choosing the amplitude of the parametric modulation field so that \(J_0\left(\left| \mu _1\right| \right)=J_2\left(\left| \mu _1\right|
\right)\), the detected signal is sensitive only to \(K_y\):
\be
\left\langle S_x\cos \left[\omega _1t+\alpha \right]\right\rangle = 
\frac{J_0J_{-1} S_z}{ \Gamma _A}\cos [\alpha ]{  }\gamma _Ab_K K_y  
\ee

The transverse polarization produces a Faraday rotation of the probe laser by an angle
\be
 \theta =n_A \sigma _0L \frac{W}{2\Delta }P_{\infty }S_x 
\ee
This equation assumes that the probe is far off resonance, \(\Delta \gg W\)\(\) where \(W\) is the linewidth of the optical transition. { }The optical
depth at line center is \(n_A \sigma _0L,\)and the circular dichroism of the probe transition is \(P_{\infty }=1 \mbox{\rm \, or} -1/2\) for D1 or D2 probe
light. { }For best signal, the detuning is chosen to moderately attenuate the probe beam, so that \(\frac{2\Delta }{W}\sim \sqrt{n_A \sigma _0L}\)
giving 
\be
 \theta \sim \sqrt{n_A \sigma _0L} P_{\infty }S_x 
\ee
The NMRG can be quite optically thick, producing Faraday rotation angles that are a radian per unit spin.  It is the large signal-to-noise ratio for this detection that allows the oscillator to have a frequency stability several orders of magnitude smaller than the resonance linewidth.

\section{Finite Gain Feedback Effects: Scale Factor and Bandwidth}

$\quad $The phase lock between the gyro phase and the feedback phase is a critical component of the NMR gyro. { }In this section we consider the
effects of finite feedback phase on the behavior of the gyro.

$\quad $The NMR phase precession is, for small deviations of the phase difference from the lock point \(\beta _0\):
\be
 \frac{d\phi }{dt}=\Omega _z+\Gamma _2(\theta -\phi ) 
 \ee
where \(\theta\) is the drive phase. { }\\

$\quad $For simplicity, we assume that the drive phase \(\theta =\phi -\beta\) is generated with simple proportional feedback of the form
\be
 \frac{d\theta }{dt}=\omega _0+g\left(\bar{\beta }-\beta _0\right)\frac{d\bar{\beta }}{dt}=\frac{-\bar{\beta }}{\tau }+\frac{(\phi -\theta )}{\tau
} \ee
For a stationary gyro, the clock-derived frequency \(\omega _0\) is tuned so that \(\omega _0=\Omega _z-\Gamma _2\beta _0.\)

\begin{figure}
\includegraphics[width=3.0 in]{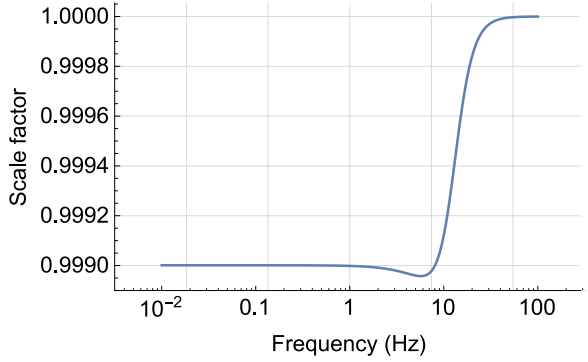}
\caption{Scale factor frequency dependence.  Parameters are $\Gamma=10$ mHz, $g=$10 Hz, $\tau=0.01$ sec.}\label{scalefactorfig}
\end{figure}

$\quad $Let us consider the response to a AC Larmor frequency \(\Omega _z=\tilde{\Omega }_ze^{i \omega  t}\). { }The corresponding frequency response
is
\be
 i \omega  \tilde{\phi }=\frac{g-\omega  (i+\tau  \omega )}{g+(\Gamma -i \omega ) (1-i \tau  \omega )}\tilde{\Omega }_z 
\ee
At low frequencies, the scale factor is \(\frac{g}{g+\Gamma }=1-\frac{\Gamma }{g+\Gamma }\) and approaches 1 at high frequencies, as shown in Fig.~\ref{scalefactorfig}.
 { }The gyro bandwidth is not limited by either $\Gamma$  {or} $g$.

\section{Noise}

$\quad $The noise characteristics of the NMRG can be understood by modifying Eq. \ref{gyroeqn} to include errors in the measurement of the relative phase of the precession and drive:
\begin{equation}
\begin{array}{c}
 \frac{d\phi }{dt}=\Omega _z-\Gamma _2\tan \left[\beta _0+\delta \beta (t)\right] \\
\end{array}
\end{equation}
Since the NMRG is supposed to accurately and precisely measure the Larmor frequency, fluctuations in \(\Omega _z\) should not be considered noise, unless
one is attempting single species gyro operation, in which case such fluctuations would constitute an unwanted background. { }Cancellation of fluctuating
magnetic fields is a primary motivation for dual species operation, and will be considered futher in Sec. \ref{sec:dual}.

$\quad $Fluctuations in the phase \(\delta \beta\) are of primary importance for noise considerations. { }We will assume that the driving fields
are noiseless, so that the dominant contribution to the phase noise is due to imperfect measurement of the NMR phase. { }According to Eq. (26), this
results in a frequency noise
\be
 \delta \tilde{\nu }(f)=\frac{\Gamma _2\delta \tilde{\, \beta }(f)}{2\pi  } 
 \ee
Under most conditions, errors in the phase measurements arise from background y-magnetic field fluctuations \(\delta \, \tilde{B}_y(f)\) leading
to a finite SNR for the detection of the Xe precession. { }Then 
\be \delta \tilde{\nu }_f(f)=\frac{\Gamma _2\delta \, \tilde{B}_y(f)}{2\pi  B_{{Xe}}} 
\ee
where \(B_{{Xe}}\) is the effective magnetic field as detected by the alkali magnetometer. In the gyro context, this is referred to as angle
random walk and is the fundamental source of rotation rate white noise.

\begin{figure}
\includegraphics[width=3.0 in]{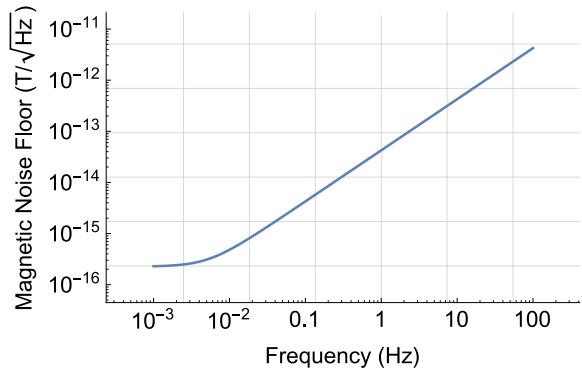}
\caption{Magnetic noise floor for an Xe-129 NMR oscillator.  Assumes $ \tilde{B}_y=0.5$ nG/$\sqrt{\rm Hz}$, $B_{Xe}=1$ mG, $T_2=30$ s. }\label{bnoiseNMR}
\end{figure}

$\quad $While the bandwidth of the NMR with feedback is quite high, the noise increases at high frequencies due to the finite SNR of the phase measurement
(angle white noise). { }This results in an effective frequency noise
\be
 \delta \tilde{\nu _{\theta }}(f)=f \delta \tilde{\beta }(f)=f\frac{\delta \, \tilde{B}_y(f)}{B_{{Xe}}} 
\ee
or an effective magnetic noise floor of
\be
 \delta \tilde{B}_z(f)&=&\frac{1}{\gamma }\sqrt{\delta \tilde{\nu _{\theta }}{}^2+\delta \tilde{\nu }_f{}^2}\nonumber \\
 &=&\frac{\delta \tilde{\, \beta }(f)}{2\pi
{  }\gamma  T_2}\sqrt{1+\left(2\pi  f T_2\right){}^2} 
\ee
as shown in Fig.~\ref{bnoiseNMR}.

$\quad $Let us now consider the statistical properties of the NMRG as a function of averaging time. At short averaging times, the finite SNR is the
limiting quantity, equivalent to angle white noise. { }We imagine passing the gyro output through a low-pass filter with time constant \(t_a\). {
}The equivalent frequency noise is
\be
 \delta \tilde{\nu _{\theta }}(f)&=&\frac{f \delta \tilde{\beta }(f)}{\sqrt{1+\left(2\pi  f t_a\right){}^2}}\nonumber \\
 &=&\frac{f}{\sqrt{1+\left(2\pi  f t_a\right){}^2}}\frac{\delta
\, \tilde{B}_y(f)}{B_{{Xe}}} 
\ee
There is a noise-bandwidth tradeoff, with the angle white noise dominating for \(t_a<T_2\). { }The figure
shows the Allan deviation \(\sigma (\tau )\) for various averaging times:
\be
\sigma ^2(\tau )&=& 2 \int _0^{\infty }df \frac{\sin  (\pi  f \tau )^4}{(\pi  f \tau )^2}\left[\delta \tilde{\nu _{\theta }}{}^2+\delta \tilde{\nu
}_f{}^2\right]\nonumber \\
&=&\left(\frac{\delta\tilde{B}_y(f)}{B_{{Xe}}}\right)^2\left(\frac{3+e^{-2 \tau \left/t_a\right.}-4 e^{-\tau \left/t_a\right.}}{16
\pi ^2 \tau ^2 t_a}+\frac{1}{8 \pi ^2\tau  T_2{}^2}\right) 
\ee

\begin{figure}
\includegraphics[width=3.0 in]{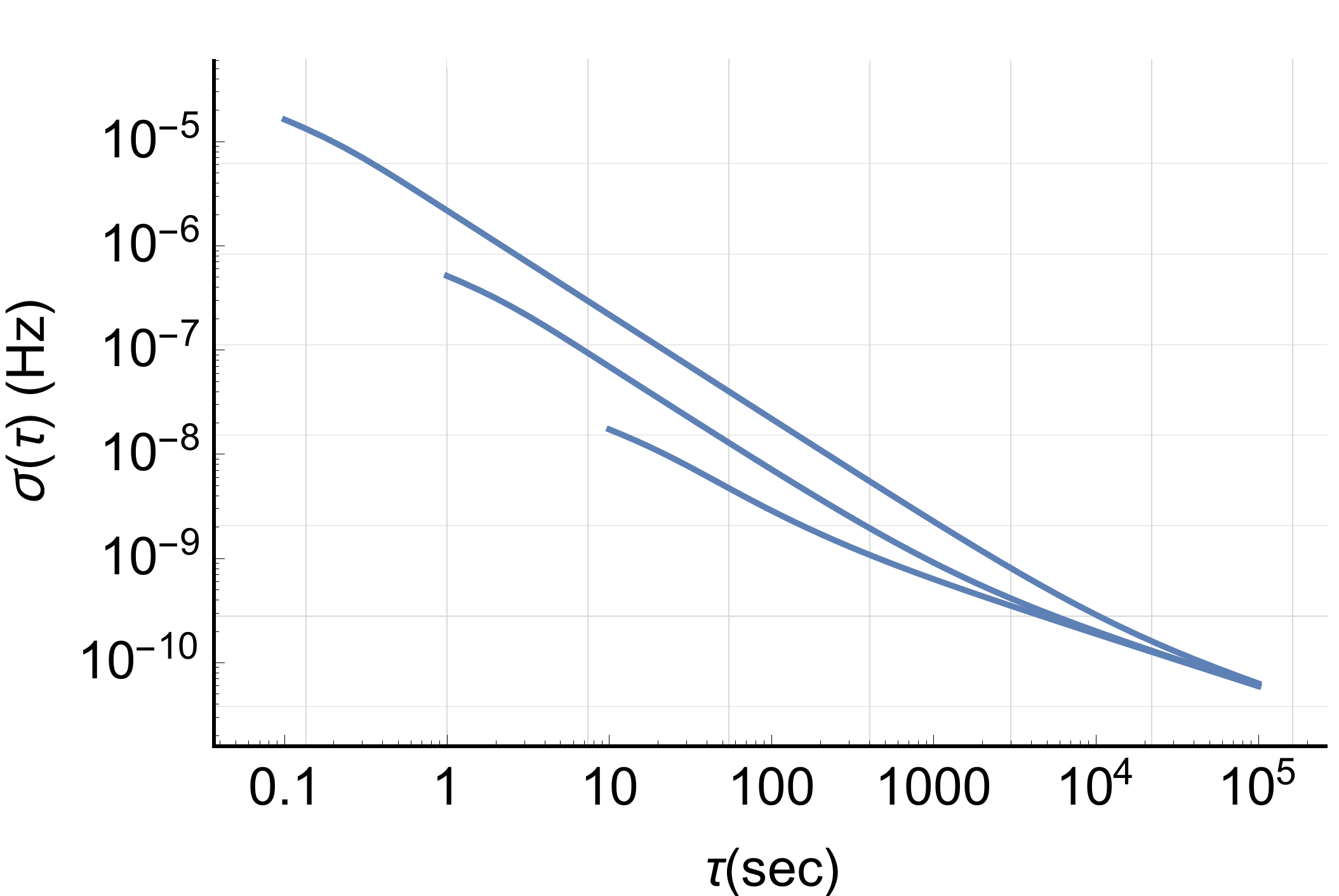}
\caption{NMR gyro rotation uncertainty (Allan deviation) as a function of time, for various values of the low-pass filter on the gyro output (0.1, 1, and 10 sec).  Fundamental noise parameters are the same as Fig.~\protect\ref{bnoiseNMR}. }
\end{figure}

   \section{Dual Species Operation} \label{sec:dual}

$\quad $Unless one is using the NMRG to measure magnetic fields, the major error encountered is from magnetic field noise. { }Thus it is key to use
two NMR species, with one being used to stabilize the magnetic field by feedback to a clock-generated reference frequency. { }Then rotations are
detected by comparison of the second isotope to a second clock-generated frequency. In this manner magnetic noise is cancelled \citep{Grover79patent}.

$\quad $We now generalize Eq. (13) to two species:
\be
 \frac{d\phi _a}{dt}&=&\gamma _aB_z+\omega _r+\Delta _a=c_a\\
 \frac{d\phi _b}{dt}&=&\gamma _bB_z+\omega _r+\Delta _b 
\ee
The offsets \(\Delta _i\) include both purposeful phase shifts between the drive and precession signals, and other sources of bias to be discussed
below. { }

There are many potential ways to implement two-species operation. { }Perhaps the simplest in concept is to feedback to the magnetic field to stabilize species $a$ to a frequency
\(c_a\) that is derived from a stable clock. { }Then compare the precession of species \(b\) to a second clock-derived frequency \(c_b:\)
\be \frac{d\phi _b}{dt}-c_b&=&\left[\frac{\gamma _b}{\gamma _a}c_a-c_b+\frac{\gamma _a\Delta _b-\gamma _b\Delta _a}{\gamma _a}\right]+\omega _r\left(1-\frac{\gamma_b}{\gamma _a}\right)\\
&=&\omega _b+\omega _r\left(1-\frac{\gamma _b}{\gamma _a}\right) \label{dualfreq}
\ee
The terms in the bracket combine to produce an over-all bias \(\omega _b\) that can be tuned to zero if desired by adjusting \(c_b\) or the individual
offsets. { }In this implementation, the rotational scale factor is \(\left(1-\frac{\gamma _b}{\gamma _a}\right)\) which is to a high degree a {``}constant{''}
of nature. { }In fact, there are known weak dependences on gas pressure, temperature, etc, but they begin to occur in the 7th decimal place \citep{Brinkmann62}. { }Note
that since \(\, ^{131}{Xe}\) has the opposite sign of the other nuclei, a dual-species NMRG that includes that isotope will have a rotational
scale factor greater than 1. 

$\quad $More about the signs: in order to avoid a proliferation of \(\pm\) symbols, our convention is that the clock frequencies \(c_{a,b}\) are taken
to have the same sign as their respective magnetic moments. { }Thus if \(\, a=^{129}{Xe}\), \(b=\, ^{131}{Xe}\), \(c_a<0\) and \(c_b>0\)
and the quantity \(\frac{\gamma _b}{\gamma _a}c_a-c_b\)will be nearly zero.

$\quad $A second approach is to stabilize the {``}difference frequency{''} \(\frac{d\phi _a}{dt}-\frac{d\phi _b}{dt}\) to
a clock derived \(c_{ab}=c_a-c_b\). { }This has the advantage that \(\frac{d\phi _a}{dt}-\frac{d\phi _b}{dt}\)is independent of \(\omega _r\)
so that the magnetic field feedback does not have to compensate for rotation at high rotation rates. { }The corresponding relation for species \(b\)
is then
\be 
\frac{d\phi _b}{dt}-c_b=\left[\frac{\gamma _bc_{ab}}{\gamma _a-\gamma _b}-c_b+\frac{\gamma _a \Delta _b-\gamma _b \Delta _a}{\gamma _a-\gamma
_b}\right]+\omega _r 
\ee
which has a rotational scale factor of 1. { }

$\quad $It is also interesting to consider how noise propagates through a two-species NMRG. { }\(B_z\) fluctuations are in principle completely suppressed
by the co-magnetometer arrangement. But \(B_y\) fluctuations that result in phase noise are indistinguishable from real magnetic field changes
and are hence compensated for by the magnetic field feedback loop. { }Such fluctuations that happen to be proportional to the ratio of gyromagnetic
ratios are effectively equivalent to a magnetic field along $z$ and will be cancelled. { }The result can be seen from Eq. (35) with fluctuating \(\Delta _i:\)
\be
 \delta \tilde{\omega }_r^2&=&\left(\frac{\left| \gamma _a\right|  \Gamma _{2\, b}}{\left| \gamma _a-\gamma _b\right| }\frac{\delta
\, \tilde{B}_y\left(f_b\right)}{ B_{{Xe},b}}\right)^2+\left(\frac{\left| \gamma _b\right| \Gamma _{2\, a} }{\left| \gamma
_a-\gamma _b\right| } \frac{\delta \, \tilde{B}_y\left(f_a\right)}{ B_{{Xe},a}}\right)^2 
\ee
This relation shows that the angle-random-walk for the small-gyromagnetic ratio species is more important than for the large $\gamma $ species.

\subsection{Systematic Errors}

While the inherent statistical properties of the NMRG are impressive,  management of systematic errors is key to the long-term stability of the device. These include the differential alkali field, shifts of electric quadrupole interactions at the cell walls, and offset drifts.  Before discussing the details of these individual contributions, we present some general considerations.

For dual-species operation, the bias frequency is, from (\ref{dualfreq}),
\be
\omega_{\rm bias}=\frac{\gamma _b}{\gamma _a}c_a-c_b+\frac{\gamma _a\Delta _b-\gamma _b\Delta _a}{\gamma _a}
\ee
The first two terms represent phase drift from the system clock, which is greatly suppressed as long as the reference frequencies are close to the NMR resonance frequencies.  We will assume that a high quality clock is used such that we can ignore this contribution.  The third term represents bias from the aforementioned effects.

Any source of bias whose changes scale proportionate to the respective gyromagnetic ratios is eliminated by dual species operation. Thus magnetic noise, even if imperfectly cancelled by the magnetic field feedback loop, is not a source of bias.

A key point to note is that by adjustment of one or both of the clock frequencies, or by setting a purposeful phase shift between the drive and the nuclear precession, the bias can be set to any value that is wished, including zero.  The bias in and of itself is usually not important, but its drift (instability) with time and temperature is key.  Let us assume that the gyromagnetic ratios are temperature independent.  Then the bias instability is
\be
\delta\omega={d \omega_{\rm bias}\over dT}\delta T=\left[{d \Delta_{b}\over  dT}-{\gamma_b\over \gamma_a} {d \Delta_{a}\over dT}\right]\delta T
\ee
where $\delta T$ is the temperature instability  of the system.  As we shall see, the systematic shifts can generally be arranged so that this factor vanishes, and second-order temperature deviations set the ultimate limit.

\subsubsection{Differential alkali field}

The contribution of the alkali field shifts to bias is
\be
\omega_1&=&\frac{\gamma _a(\gamma_b B_b)-\gamma _b(\gamma_a B_a)}{\gamma _a}=\gamma_b(B_b- B_a)\\
&=&{\kappa_b-\kappa_a\over \kappa_b}\gamma_b b_{bS}S_z
\ee
When two different chemical species (He and Xe, for example) are used, this shift is comparable in size to the shift of the species with the largest $\kappa$ (Xe in this case).  This problem was recognized early in the Litton program \citep{Grover79patent} and motivates the use of two Xe isotopes where the ``isotope shift" should be very small.  The fractional isotope shift ${(\kappa_b-\kappa_a)/\kappa_a}$ was recently measured in Ref. \citep{Bulatowicz2013} to be 0.0017.  This gives a typical size of the alkali field bias to be 115 $\mu$Hz for fully polarized Rb at $10^{13}$/cm$^3$.  The temperature dependence, assuming the Rb vapor pressure variation is the dominant contributor, is roughly 7 $\mu$Hz/K=$9^\circ$/hr-K.

\subsubsection{Quadrupole shifts}

The down-side of using the two Xe isotopes is that the spin-3/2 \Xe{131} nucleus experiences electric quadrupole interactions from electric field gradients at the cell walls \citep{Kwon1981}.  The size of the quadrupole interaction can vary by an order of magnitude or more from cell to cell.  Because NMR gyros are continuously driven, the signals reach a steady-state oscillation from which the presence of a quadrupole interaction can be difficult to ascertain, since the primary effect of the quadrupole interaction is a phase shift of the precession phase as compared to the drive.  It is much more apparent in a free-induction decay \citep{Bulatowicz2013}.

\begin{figure}[htbp]
   \centering
   \includegraphics[width=3.0 in]{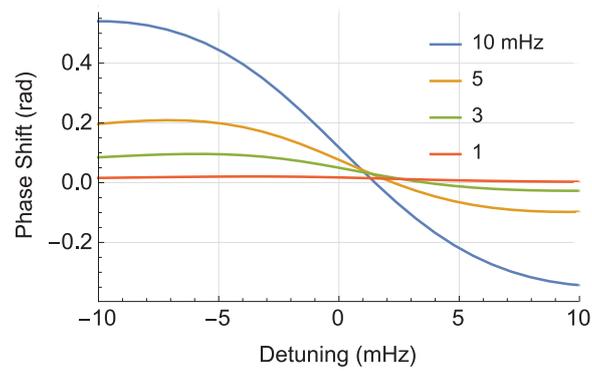} 
   \caption{Calculation of the quadrupole phase shift vs detuning for a \Xe{131} oscillator, for various quadrupole interaction strengths.  The assumed parameters are:  $T_1=T_2=20$ sec, $\Omega_d=1/T_1$, $\Gamma_{SE}=1/200$ sec.  The effective frequency shift is the NMR linewidth multiplied by the tangent of the quadrupole phase shift.}
   \label{QuadPlot}
\end{figure}

We have performed, using the methods of \citet{HJW}, a basic simulation of the first order effect of quadrupole interactions on a \Xe{131} oscillator.  Figure~\ref{QuadPlot} shows how the quadrupole contribution to the phase shift depends on detuning, for various assumed quadrupole interaction strengths.  It is interesting to note that near but not at line center the quadrupole-induced phase shift becomes relatively insensitive to the interaction strength.  This is likely closely related to the removal of transient quadrupole beats by appropriately setting the angle of the magnetic field in the rotating reference frame \citep{Wu90}.

\subsubsection{Offset Bias}

Minimization of bias instability normally favors running the two oscillators off resonance, so that there is a non-zero phase shift $\beta_0$ between the drive and precession phases.  This produces a purposeful frequency shift
\be
\omega_{\rm off}=\Gamma_2\tan\beta_0
\ee
Due to the temperature/density sensitivity of $\Gamma_2$, especially for Xe-129,  this can be a source of bias instability.  For dual species operation, the offset bias is
\be
\omega_{\rm bias}=\Gamma_{2,b}\tan\beta_{0,b}-{\gamma_b\over \gamma_a}\Gamma_{2,a}\tan\beta_{0,a}
\ee
A very important point to note is that $\omega_{\rm bias}$ is a signed quantity of essentially arbitrary magnitude (though it is impractical to operate the oscillators at more than a few linewidths off resonance).  Assuming that the Xe-129 linewidth is proportional to Rb vapor pressure and dominates the offset bias gives a typical temperature sensitivity of 100 $\tan\beta_0$ $\mu$Hz/K=150$^\circ$ $\tan\beta_0$ /hr-K.

\subsubsection{Bias Instability Compensation}

For purposes of gyro operation, a fixed bias or even trend (steady rate-of-change of bias) is acceptable.  However, uncontrolled non-magnetic bias drifts (those that do not scale with the gyromagnetic ratios) are generally indistinguishable from actual rotations and represent the ultimate precision measurable by the NMR gyro.  The most likely source of bias drifts is imperfect temperature stabilization, though pump laser intensity variation may also be a significant contributor.  Assuming that temperature variations (which may couple to pump laser intensity variations for compact systems in which the lasers are located close to the heated cell) dominate, the bias instability is approximately
\be
\delta\omega_1+\delta\omega_Q+\delta\omega_{\rm bias}=\delta\omega_1+\delta\omega_Q-{\gamma_b\over \gamma_a}\delta\Gamma_{2,a}\tan\beta_{0,a}
\ee
where we have assumed that the Xe-129 offset dominates the temperature sensitivity of the offset bias.  The key point is that the linear dependence of bias on temperature vanishes when
\be
\tan\beta_{0,a}={\gamma_a\over \gamma_b}{\delta\omega_1+\delta\omega_Q \over \delta\Gamma_{2,a}}=0.05
\ee
where the numerical value is an estimate assuming the differential alkali field is the dominant contributor to bias drifts.  Thus a modest offset of the Xe-129 frequency from resonance can eliminate the first order contributions to bias instability.

To the extent that both the differential alkali field and the offset bias are proportional to [Rb], the bias sensitivity to temperature would be cancelled to all orders.  As this assumption is likely violated at some level that may be quite implementation dependent, we note however that the Xe-131 offset can also be used to cancel second order dependencies.  Even if that cannot be done, a suppression of a factor of 100 of the bias sensitivity would produce a bias instability with 10 mK temperature stabilization of
\be
\delta\omega={7 \mu{\rm Hz/K}\over 100 }\times   .01 {\rm K}=0.7 {\rm nHz}=9 \times 10^{-4}{\rm deg/hr}.
\ee
With a achievable 10 mK temperature stability, this implies that the NMR gyro has a remarkable potential bias stability.

\section{The Northrop Grumman gyro}

\begin{figure}[htbp]
   \centering
   \includegraphics[width=3.0 in]{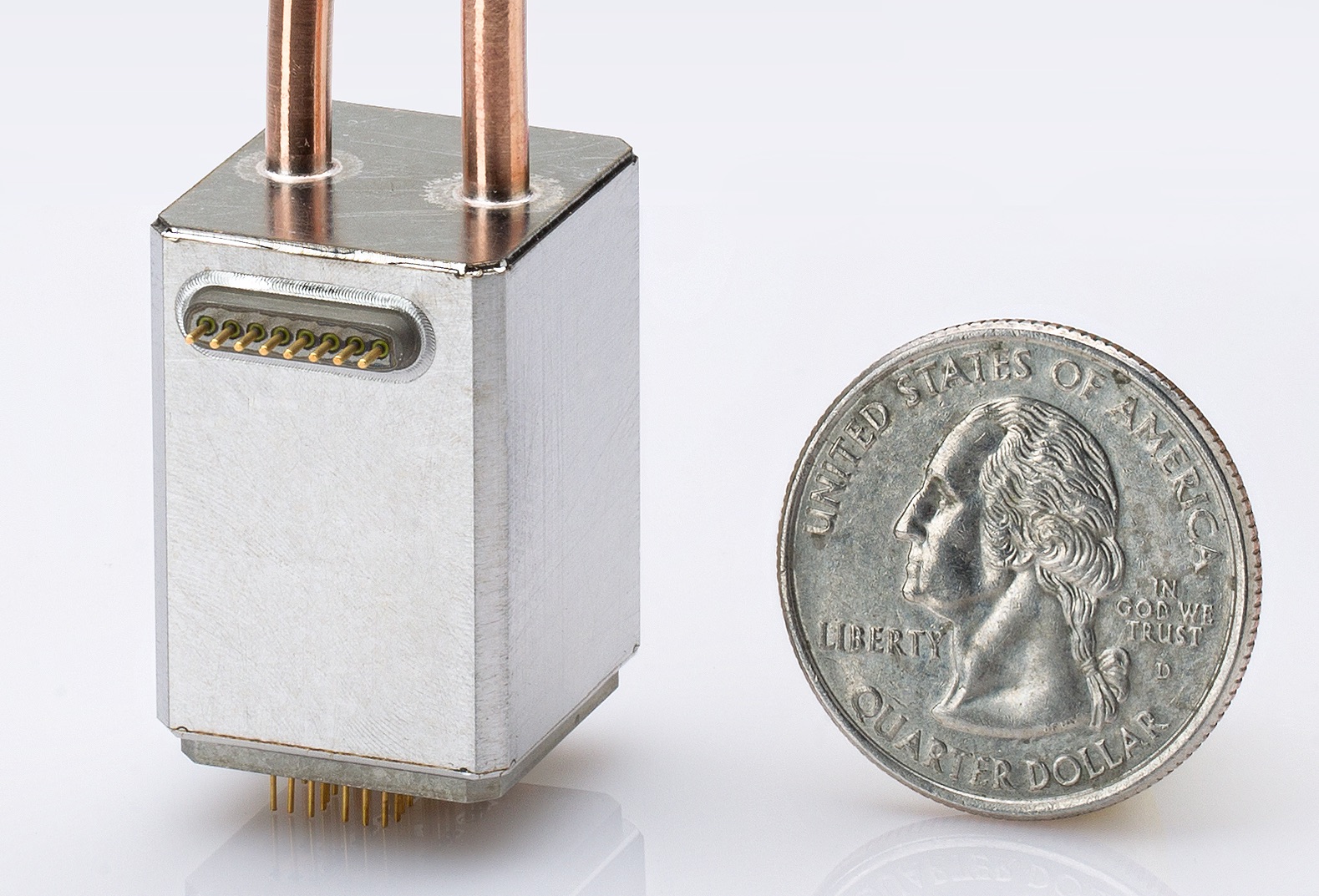} 
   \caption{Phase 4 NGC NMR Gyro physics package.  The lasers, field coils, cell with heaters, and optics are all contained within the evacuated magnetic shield.  The headers connect the physics package to the external electronics.}
   \label{NGCphysics}
\end{figure}

\begin{figure}[htbp]
   \centering
   \includegraphics[width=3.0 in]{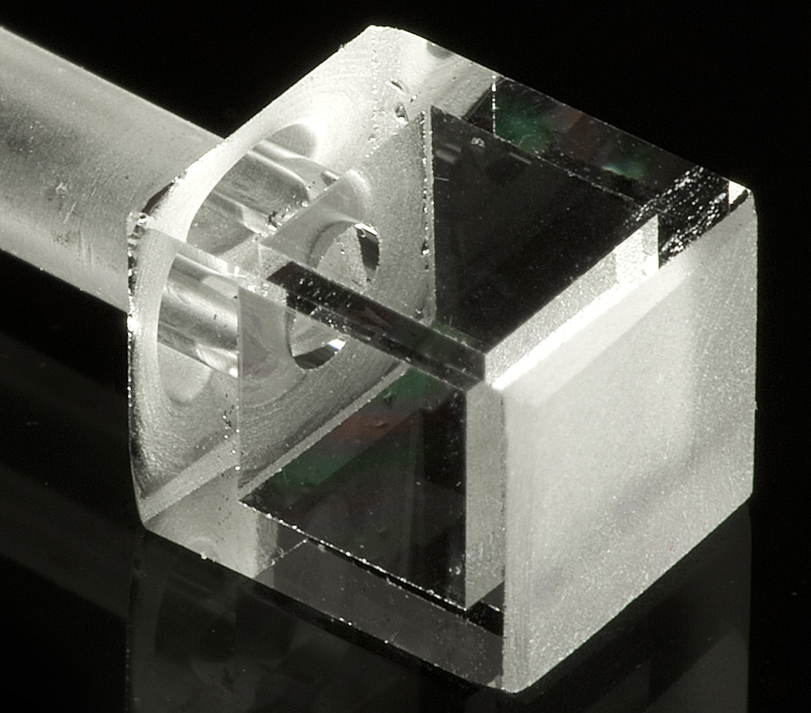} 
   \caption{Glass cell with holder.}
   \label{fig:glass}
\end{figure}

This section is a brief overview of the NMR gyro as developed over the last few years at Northrop-Grumman.  Figure~\ref{NGCphysics} is a photograph of a recent version of the gyro.  The case is a hermitically sealed and evaculated magnetic shield, and contains all the gyro components except the electronics.  The heart of the apparatus is a mm-scale cubic glass cell (Fig.~\ref{fig:glass}) containing Rb metal, isotopically enriched Xe, nitrogen buffer gas, and a small amount of hydrogen gas that forms an Rb-H coating that is known to give long 131-Xe lifetimes \citep{Kwon1981}.  The cell is held by a low thermal conductivity mount, and heated with non-resonant AC current heaters designed to minimize stray magnetic field fields from the heaters.  The vacuum, maintained by a getter pump, holds the thermal load to tens of mW at the typical $>120^\circ$C operating temperature.  Inside the shield are also a variety of magnetic field coils for providing the Gauss-level  bias field, the parametric modulation field for the alkali magnetometer, and shimming fields to optimize the transverse relaxation times of the noble gas.

Two VCSEL lasers provide up to 2.5 mW of power each for pumping and probing.  Each VCSEL is temperature and current controlled to allow selection of optimum power and tuning parameters.  An integrated optical system delivers the laser light to the cell.  The probe laser is detected by a balanced Faraday detector.

A very important component of an NMR gyro system is the electronics for control and measurement.  As there are many design choices to be made, we will content ourselves here with an overview.  A high quality quartz oscillator provides the reference clock for the system.  From it are derived the parametric modulation waveform, and reference waveforms for the two isotopes.  The Xe precession as detected by the Rb magnetometer is Fourier analyzed into separate waveforms for the two isotopes, which are amplified and phase-shifted to provide the drive waveforms for the NMR.  The two Xe waveforms are mixed to compare to the difference frequency, and a feedback loop adjusts the magnetic field to lock the difference frequency to the reference waveform from the clock.  The phase difference between the 131-Xe signal and another clock-derived reference frequency then gives a direct readout of the rotation angle.

\begin{figure}[htbp]
   \centering
   \includegraphics[width=4.0 in]{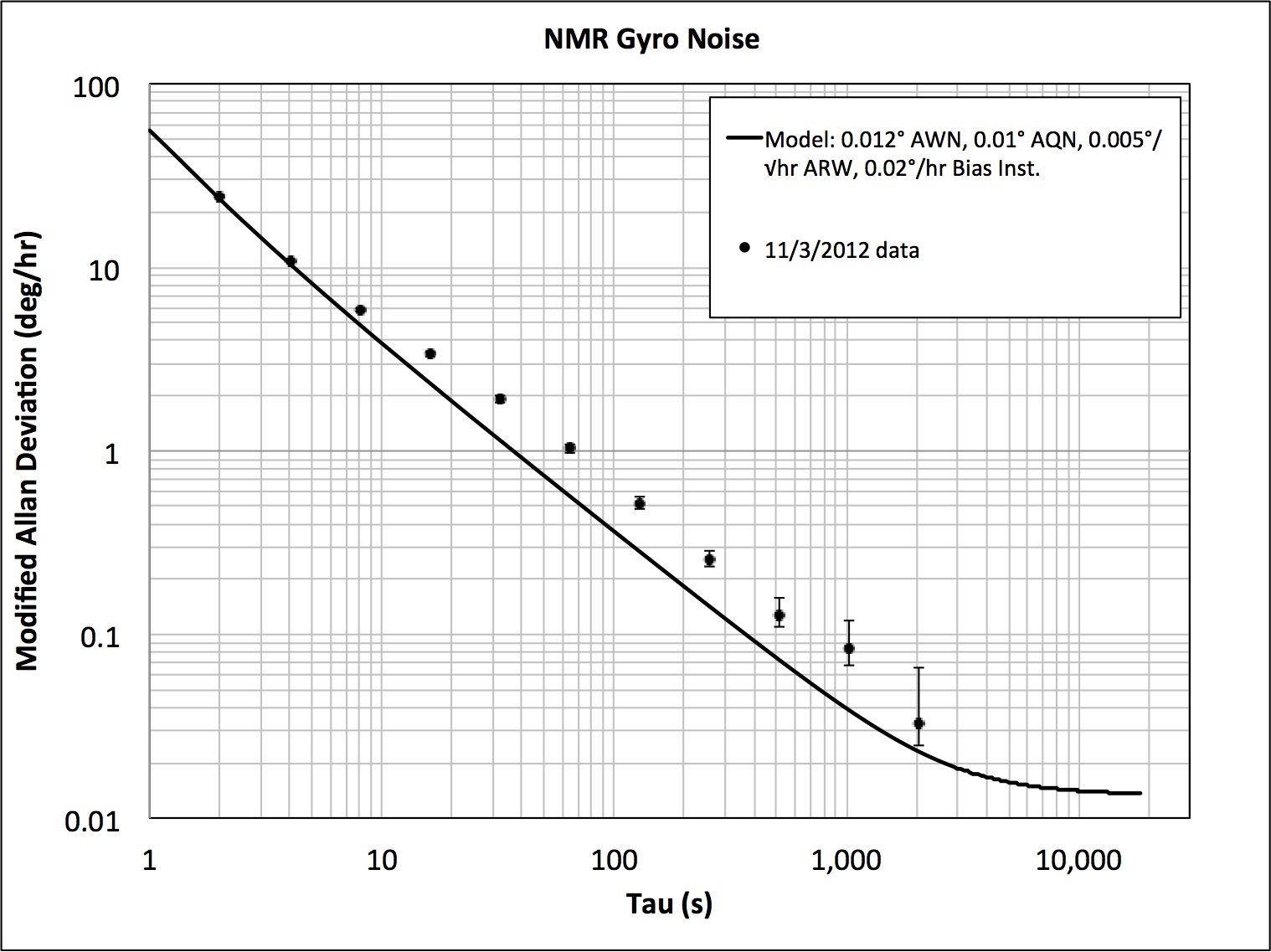} 
   \caption{Gyro noise measurement vs averaging time $\tau$.  The solid line shows a $\tau^{-3/2}$ dependence consistent with angle white noise, out until long times where the bias stability takes over. Adapted from \cite{Meyer2014}.}
   \label{fig:ARW}
\end{figure}

Table~\ref{results} summarizes performance as of 2014.  The angle-random-walk measurement of $0.005$ deg/$\sqrt{\rm hr}$ (230 nHz/$\sqrt{\rm Hz}$) is an upper limit as the system appeared to be limited by white phase noise (Fig.~\ref{fig:ARW}) until it hit its bias stability limit of 0.02 deg/hr (15 nHz).  Of course, in a practical gyro many other parameters are of importance.  One of particular interest is that the scale factor, set by the physics of the device and not any geometrical factors, is within unity to very high precision, and is tremendously stable (4 ppm turn-on to turn-on, 1 ppm over 1 day continuous operationsd).  Likewise, the full scale rate and bandwidth are  high, greatly exceeding the inherent 10 mHz bandwidth of the Xe nuclei.  Of course, as explained previously, this is due to the active feedback in the oscillator configuration but it should be noted that the 300 Hz bandwidth is 30000 Xe line-widths.  Similarly, the tight locking of the magnetic field to the difference frequency allows magnetic field suppression by a factor of 10 billion.

 \begin{table}
 \caption{NGC NMRG performance metrics, as of 2014 \label{results}}
 \begin{center}
 \begin{tabular}{l|l|l}
Metric&Unit&Performance \\ \hline\hline
Angle Random Walk& deg/$\sqrt{\rm hr}$& 0.005\\ \hline
Bias Drift &deg/hr& 0.02\\ \hline
Scale Factor&  &$0.998592(4)$\\ \hline
Scale Factor Stability& ppm& 4 \\ \hline
Full scale rate & deg/sec & 3500 \\ \hline
Bandwidth & Hz& 300 \\ \hline
Size & cm$^3$ & 10 \\ \hline
B-field suppression & & $>10^{10}$ \\ \hline
 \end{tabular}
 \end{center}
 \end{table}

Finally, we remark on Fig.~\ref{NGCphysics}, showing that these performance metrics are achieved in a very small volume.  Other gyro technologies such as ring-laser-gyros and atom inteferometers have achieved better noise characteristics but in much larger volumes than the 10 cc shown.  As the current NMR gyro demonstrations seem to be limited by technical noise, there is tremendous potential to improve on ARW.  Control of bias drift is likewise a topic of great interest and intense study.

\section{Outlook}

In this paper we have summarized the basic physics behind the operation of spin-exchange pumped NMR gyros.  Beyond the specific applied physics problem of high precision measurement of rotation in a small package, the NMR gyro represents a basic spectroscopic tool that could contribute to studies of fundamental symmetries such as searches for exotic particles, violation of local Lorentz invariance, and setting limits on permanent electric dipole moments.

We are optimistic that further development work and improved engineering of the NMR gyro will lead to improvements in ARW and bias stability while maintaining the very impressive performance metrics of bandwidth, scale factor stability, etc. that are of great importance for practical implementation of the gyro, and have been the focus of recent NGC efforts.  In particular we note that polarization and magnetic field reversals might be used, as was done by \cite{Bulatowicz2013}, to actively measure and compensate for alkali field and quadrupole shifts.
We note that fundamental noise limits have not yet been reached.  It would be very interesting to see what noise performance could be attained with an NMR gyro system optimized solely for noise performance.  Such a system might feature larger volumes, allowing for narrower line-widths and corresponding reduction of alkali and quadrupole fields.  It would also likely include new techniques for addressing sources of bias.

In the past few years, we have become interested in a new approach to NMR gyros with the potential to eliminate alkali field shifts and/or quadrupole shifts.  The basic concept is to cause the noble-gas and alkali atoms to co-precess with purely transverse polarizations.  Since the alkali polarization would be transverse to the bias field, there would be no DC alkali field parallel to the bias magnetic field, eliminating this extremely important source of bias drift.  However, since the alkali atoms have much larger magnetic moments than the Xe, enabling the transverse co-precession requires effective nulling of the alkali magnetic moment.  This is accomplished by replacing the DC bias field by a sequence of short ($\mu$s scale) alkali 2$\pi$ pulses \citep{Korver2013}, so that in a time-averaged manner the alkali atoms do not precess.  This allows for synchronous pumping of the alkali and Xe atoms at the Xe resonance frequency \cite{Korver2015}, all the while keeping the favorable fundamental statistical noise properties inherent to spin-exchange-pumped NMR gyros.

Preparation of this paper by T.W. was supported in part by the National Science Foundation (GOALI PHY1306880) and Northrop-Grumman Corporation.  This paper describes work pioneered by many Litton and NGC employees, with recent developments in particular from Robert Griffith and Phil Clarke (electronics), Michael Bulatowicz (mechanical and system design), and James Pavell (cells).

\appendix

\section{\hspace{0.7 in}RbXe Spin Exchange Rates}\label{RbXe}

In the short molecular lifetime ($\tau$) limit \citep[Eq. 109]{Happer84} spin-exchange interactions in bound and quasi-bound van der Waals molecules polarize the Xe nuclei at the rate
\be
\frac{d\expect{K_z}}{d t}=\frac{1}{T_X}\left(\frac{\alpha  \tau }{[I]\hbar }\right)^2\left[\left\langle K^2-K_z^2\right\rangle \left\langle F_z\right\rangle -\left\langle F^2-F_z^2\right\rangle \left\langle K_z\right\rangle \right]
\ee
Here the molecular formation rate is $1/T_X$, the molecular lifetime is $\tau$, the alkali nuclear spin is $\pmb I$ and the alkali total spin is $\pmb F=\pmb I+\pmb S$. 
In the very short lifetime limit  \citep[Eq. 121]{Happer84} the electron spin is decoupled from the alkali nucleus and this changes to
\be\frac{d\expect{K_z}}{d t}=\frac{1}{T_X}\left(\frac{\alpha  \tau }{\hbar }\right)^2\left[\left\langle K^2-K_z^2\right\rangle \left\langle S_z\right\rangle -\frac{1}{2}\left\langle K_z\right\rangle \right]
\ee
Binary collisions obey the same rate equation as the very short lifetime molecules, but are independent of the molecular formation and breakup times:
\be
\frac{dK_z}{d t}=\Gamma _{\rm bin}\left[2\left\langle K^2-K_z^2\right\rangle \left\langle S_z\right\rangle -\left\langle K_z\right\rangle \right]
\ee
The transition from short to very short collisions is accounted for by the factor $J=(1+\omega_{hf}^2 \tau^2)^{-1}$, which is the fraction of molecules that are broken up before precessing by a hyperfine period $2\pi/\omega_{\rm hf}$.

We define the spin-exchange rate to be the spin-exchange contribution to $T_1$ for the Xe nuclei:
\be
\Gamma _{{SE}}=\Gamma _{\rm bin}+\frac{1}{2T_X}\left(\frac{\alpha  \tau }{\hbar }\right)^2\left(J+(1-J)\frac{2\left\langle F^2-F_z^2\right\rangle }{[I]^2}\right)
\ee

We can give an explicit formula for  $\left\langle F^2-F_ z^2\right\rangle$ if we assume that the alkali spins are in spin-temperature equilibrium, $\rho=e^{\beta F_z}$.  Written in terms of the electron polarization $P=\tanh[\beta/2]$,
\be
2\expect{F^2-F_ z^2}=q={\expect{F_z}\over \expect{S_z}}=\left(\frac{8}{3 P^2+1}+\frac{8}{P^2+3}+2\right) 
\ee
where the right-hand side is specific to $^{85}$Rb with $I=5/2$. Then
\be
\Gamma _{{SE}}=\Gamma _{\rm bin}+\frac{1}{2}\frac{1}{T_X}\left(\frac{\alpha  \tau }{\hbar }\right)^2\frac{1+q (\omega  \tau )^2/[I]^2}{ 1+(\omega  \tau )^2}
\ee
should accurately represent the spin-exchange rate as long as the total pressure exceeds a few tens of Torr.

Detailed balance allows the molecular formation time to be rewritten in terms of the molecular breakup time, the alkali density, and the chemical equilibrium coefficient $k_{\rm chem}$.
\be
\frac{[X]}{T_X}=\frac{[{AX}]}{\tau }=\frac{k_{{\rm chem}}[A][X]}{\tau }\to \frac{1}{T_X}=\frac{k_{{\rm chem}}[A]}{\tau }
\ee

It is beyond the scope of this paper to review the often conflicting literature on RbXe spin-exchange measurements, but we have generally found the following numbers to give reliable estimates in our experiments.  For a He-dominated buffer gas, 
\citet{Nelson01b} measured $\omega\tau_{\rm He}=2.95$ amagat/[He], and $k_{chem}=$213 \AA$^3$ at 80$^\circ$C, somewhat smaller at 120.
 \citet{Ramsey83} showed that $\tau_{N_2}=\tau_{He}/1.6$. 
\citet{Bhaskar82} deduced  $\gamma N/\alpha=4.1$ for Xe-129, and  
\citet{Bhaskar83} measured $\gamma N/h=$120 MHz from magnetic decoupling measurements, so $\alpha/h=$29 MHz.  The binary collision contribution to the spin-exchange rate was measured by \citet{Jau02} to be $\Gamma _{\rm bin}/[{\rm Rb}]=1.75\times 10^{13}$ cm$^3$/s.

\section*{References}

\bibliographystyle{elsarticle-harv}

\bibliography{spinexchange}

\end{document}